\documentclass[preprint]{aastex}
\usepackage{natbib}
\shorttitle{TWA Stellar Parallaxes}
\shortauthors{Weinberger et al.}

\begin{document}
\renewcommand{\topfraction}{1.0}
\renewcommand{\bottomfraction}{1.0}
\renewcommand{\floatpagefraction}{0.7}
\renewcommand{\textfraction}{0.2}

\title{Distance and Kinematics of the TW Hydrae Association from Parallaxes}

\author{Alycia J. Weinberger}
\affil{Department of Terrestrial Magnetism, Carnegie Institution of
Washington}
\affil{5241 Broad Branch Road NW, Washington, DC 20015}
\email{weinberger@dtm.ciw.edu}

\author{Guillem Anglada-Escud\'{e}}
\affil{Universit\"at G\"ottingen, Institut f\"ur Astrophysik}
\affil{Friedrich-Hund-Platz 1, 37077 G\"ottingen, Germany}

\author{Alan P. Boss}
\affil{Department of Terrestrial Magnetism, Carnegie Institution of
Washington}
\affil{5241 Broad Branch Road NW, Washington, DC 20015}

\begin{abstract}

From common proper motion and signatures of youth, researchers have identified
about 30 members of a putative TW Hydrae Association. Only four of these had
parallactic distances from Hipparcos. We have measured parallaxes and proper
motions for 14 primary members. We combine these with literature values of
radial velocities to show that the Galactic space motions of the stars, with
the exception of TWA 9 and 22, are parallel and do not indicate convergence at
a common formation point sometime in the last few million years. The
space motions of TWA 9 and 22 do not agree with the others and indicate that they
are not TWA members.  The median parallax is 18 mas or 56 pc.  We further
analyze the stars' absolute magnitudes on pre-main sequence evolutionary
tracks and find a range of ages with a median of 10.1 Myr and no correlation
between age and Galactic location. The TWA stars may have formed from an
extended and filamentary molecular cloud but are not necessarily precisely
coeval.
\end{abstract}


\section{Introduction}

Nearby young stars of age $\sim$5--10 Myr provide our best opportunity to
study the late stages of star and planet formation with high sensitivity and
spatial resolution. During this time period, the last gas-rich disks
dissipate, and the onset of the debris disk phase occurs.  The star TW Hydrae
sports a massive disk and was first identified as an isolated T Tauri star
\citep{Rucinski:1983} and then the first member of an association of young
stars \citep{delareza:1989,Gregoriohetem:1992,Kastner:1997}. As the nearest
example of a protoplanetary disk, TW Hya has been studied at every available
wavelength and spatial resolution, but to understand its disk in context, TW
Hya's age must be well-determined and its disk evolution compared to other
stars of similar age and mass.

Over the last decade, $\sim$30 members of a putative TW Hydrae Association
(TWA) have been identified from a combination of stellar-activity searches and
space motion studies
\citep[e.g.][]{Webb:1999,Sterzik:1999,Zuckerman:2001,Gizis:2002,Song:2003,Scholz:2005,Looper:2007,Shkolnik:2011}. Members
range in spectral type from A0 to brown dwarfs and have presumed common ages
$\sim$10 Myr, as determined from Li depletion and modeling on pre-main
sequence evolutionary tracks \citep{BarradoYNavascues:2006}. Disks in TWA
range from the four accreting, gas rich, protoplanetary ones (TWA 1, 3, 27,
and 30), to seven transitional or debris disks (TWA 4B, 7, 11A, 26, 28, 31 and
32), to the majority of members that have no detectable disks at all
\citep{Weinberger:2004,Low:2005,Riaz:2008,Plavchan:2009,Schneider:2012}.

The Hipparcos mission determined distances to only four TWA members, including
TW Hya itself, for an average distance of 70 pc. Although the association
contains the youngest stars close to the Sun, most of its members are low-mass
stars and were therefore too faint for Hipparcos. Ground-based parallaxes have
been measured for two brown dwarfs in the association -- TWA 27 (2M1207)
\citep{Gizis:2007} and TWA 28 (SSSPM J1102) \citep{Teixeira:2008}, and one
additional star that may not actually be a member -- TWA 22AB,
\citep{Teixeira:2009}.  This highlights the problem of determining association
membership -- although their proper motions are similar, as ensured by search
methods, without distances and therefore true Galactic space motions TWA stars
can only be presumed to be associated.  Hence, various members have been
suggested and then excluded by proper motion studies using varyingly
restrictive conditions for membership
\citep[e.g.][]{Makarov:2001,Song:2003,Mamajek:2005}.

Distances to a substantial number of members would aid in defining the
association and estimating its age from pre-main sequence tracks. If the
kinematics allow, a cluster expansion age could also be determined. In this
paper we present parallaxes and proper motions to 14 primary stars and two
visual binary companions.

\section{Data}

We have observed 14 TWA primary members with the CAPSCam instrument at the
2.5m DuPont Telescope at Las Campanas Observatory.  The instrument and basic
data reduction techniques are described in \citet{Boss:2009} and
\citet{Anglada:2012} and only briefly summarized here.  CAPSCam utilizes a
Hawaii-2RG HyViSI detector filtered to a bandpass of 100 nm centered at 865 nm
with 2048x2048 pixels, each subtending 0\farcs196
on a side.  The main advantage of this camera is its ability to achieve
simultaneously high S/N on a bright target star and fainter astrometric
reference stars.  Our typical TWA sources have I$\approx$10 (see Table
\ref{tab_obs}); we place these bright target stars in an independently
readable subarray called the ``guide window'' (GW). We typically locate the GW
in the center of the full field (FF) and rapidly read and reset it for
integration times down to 0.2~s for a 64$\times$64 pixel
subarray. Simultaneously, the FF integrates for as long as necessary to obtain
high S/N on many reference stars. Table \ref{tab_obs} gives typical
integration times for each target in the GW and FF, although the times were
adjusted during each epoch to account for seeing and clouds.  

We drew our sample from stars without parallaxes (thus excluding TWA 1, 4AB,
9AB, 11A, 22, 27, and 28). We only observed bona fide members identified in
the convergent point analysis of \citet{Mamajek:2005}; thus we excluded TWA
17, 18, 19AB, and 24. Although \citet{Mamajek:2005} exclude TWA 12, we
included it because of its large discrepancy between photometric
\citep{Song:2003} and kinematic distance. We excluded TWA 3AB for
being too bright. CAPSCam saturates in 0.2 s on I$\approx$9. We excluded TWA
6, 7, and 11B for being near very bright stars that would fall within the
CAPSCam field of view because saturated images leave long-lasting after-images
on the detector. We excluded TWA 8AB because the two stars, separated by
13\arcsec\ do not fit in the 64x64 GW.  Three targets -- 30AB, 31, and 32,
were discovered after our survey began \citep{Looper:2010, Shkolnik:2011}.
The remaining stars were all observed with the exception of TWA 10 (Table
\ref{tab_obs}).  CAPSCam could in the future provide parallaxes for the
remaining stars with I$>$9 -- TWA 8AB, 10, 17, 18, 30, 31 and 32.

We can operate in a mode, called ``Guide-window shutter'' (GWS) where an iris
shutter opens over the FF only when the GW is integrating and closes during
the initial pixel resets and during each GW read. This is done to guarantee
that the images in the GW and FF sample the atmosphere identically and that no
astrometric bias is introduced between them. Some of our data were taken with
the shutter in this mode, but some were taken with no shutter (NS) and some in
a mode where the shutter closes during the initial pixel resets but remains
open during each GW read (NGWS).  In principle, the best astrometric precision
is obtained in GWS mode, but we operated without it in order to improve
efficiency.

Flat field images to correct for pixel-to-pixel variations either intrinsic to
the detector or due to the finite opening time of the iris shutter are
obtained while exposing on a quartz lamp projected on a flat-field screen.

The imaging of each astrometric field is repeated typically 15-20 times at
each epoch.  For each FF image, we obtain many GW images. In post-processing,
these are summed and and re-inserted into the FF image.
For each night, one image is selected as an astrometric template of the field
for that night.  Sources are found automatically, and a fine centroiding
algorithm is applied to produce a catalog with the sub-pixel positions of all
of the objects in all of the images of the field for a given night.

Each star was observed in at least four independent epochs, with all but one
star observed in five or more, i.e. dates separated by enough days that they
provide independent constraints on the parallax. The dates of observation are
given in Table \ref{tab_obs}.  Data from all epochs are combined in an
astrometric solution to derive the positions, proper motions, and parallaxes
of all the stars in each target field.  The astrometric solution is an
iterative process. An initial catalog of positions is generated from a chosen
epoch, a transformation is applied to every other epoch's catalog to match the
initial catalog, and the apparent trajectory of each star is then fitted to a
basic astrometric model. The initial catalog is updated with new positions,
proper motions, and parallaxes, and a subset of well-behaved stars is selected
to be used as the reference frame.  The reference stars must be successfully
extracted in every epoch and a subset of at least 15, and more typically 30,
are chosen that show the smallest epoch-to-epoch variation in their solutions.
Over all of our target fields, these stars have typical I-band magnitudes of
13.5 -- 17.9 with a median of 16.2.  This process is then iterated a small
number of times. Again, details may be found in \citet{Anglada:2012}.

\subsection{Zero-Point Parallax Correction \label{sec_zpt}}

The motion of the target star is measured with respect to background stars, which are not
truly stationary and that all have parallactic motions given by the Earth's motion and
therefore move in the same direction.  This introduces a small bias, so the average
parallax of the reference stars must be removed to find the absolute parallax.

If all the stars in the field had perfectly known distances and that information was
inserted a priori, the parallax zero-point correction would always be a positive
number. Note, however, that some zero-point corrections in Table \ref{tab_astrometry} are
negative. Because we do not know the distances a priori, in the astrometric solution, the
parallaxes for all objects are initialized to zero. In the iterations that follow, each
individual parallax and proper motion is adjusted. While the mean parallax measurement
over all the stars after the first iteration should still be approximately zero, the mean
parallax of a subset of them, i.e. those used as reference stars, can be either positive or
negative due to statistical fluctuations.  At any epoch, the position of a star has
centroiding uncertainties, and for distant stars, proper motion will take out all apparent
motion of the star leaving positional residuals that are both positive and
negative. Therefore, although the true parallax to every star must be positive, we allow
the fit parallaxes to take on positive and negative values.  The quality of the final
astrometric solution as measured by the residuals on all the stars is independent of the
value of the mean zero-point parallax and will not be adjusted in subsequent iterations.

To find the zero-point for each field, we estimate a photometric distance to
the brightest reference stars by fitting a Kurucz stellar model to catalogued
USNO-B1 -- B2, R2, and I; \citep{Monet:2003} and 2MASS -- J, H, and K$_s$
\citep{Skrutskie:2006} photometry and assuming each star is a dwarf. Giant
stars are then easily recognizable because they appear to be so close that
they should have detectable parallaxes, and we refit them as giants.  Dwarf
stars with fit T$_{\rm eff}<$3800~K are not considered because the stellar models
are less reliable. We average the difference between our astrometrically
determined (even if they are not statistically significant) and photometric
parallaxes to find the average bias and its uncertainty and subtract it from
our relative parallaxes and propagate the uncertainty. A comparison of our
parallaxes determined this way to literature values is given in
\citet{Anglada:2012}.

\begin{figure}
\includegraphics[scale=0.8]{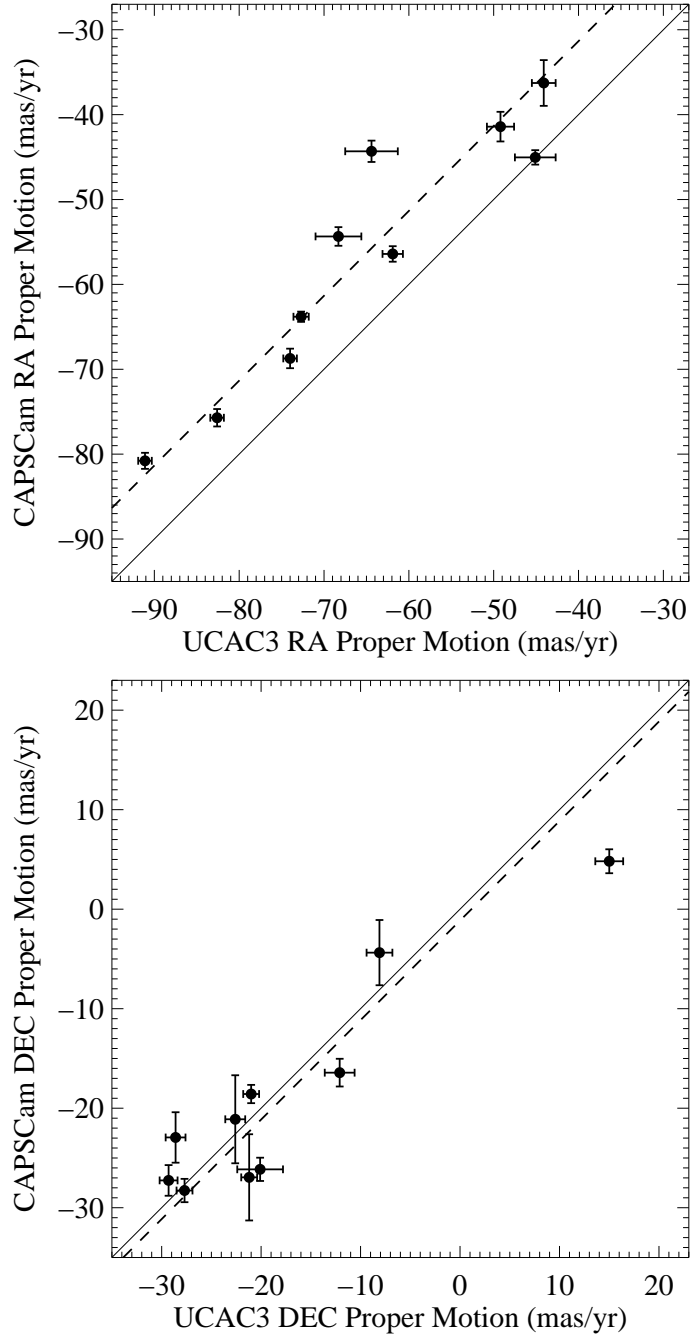}
\figcaption{Comparison of proper motions of the ten TWA stars both reported in UCAC3
  \citep{Zacharias:2009} and measured in our survey. The dashed lines show the
  mean offsets of 8.8 mas/yr in RA and -1.1 mas/yr in DEC between our relatve
  proper motions and the absolute UCAC3 ones. \label{fig_compare_ucac}}
\end{figure}

In principle, this zero-point correction could be done for the proper motions
as well, but the reference stars do not generally have catalogued proper
motions to use in measuring their bias. Instead, we estimate our bias directly
by comparing the proper motions of the TWA stars themselves as computed from
our astrometry with their cataloged values in UCAC3 \citep{Zacharias:2009}.
Given that the brightnesses and spectral types of the reference stars have
approximately the same distribution for all our targets, which are also in the
same general direction in the Galaxy, we can then find a correction of our
CAPSCam proper motions to absolute proper motions. Leaving out TWA 13 and 15,
which are visual binary stars whose proper motions in the UCAC3 are suspect,
ten of our 14 stars have UCAC3 proper motions. CAPSCam-determined proper
motions are indeed well correlated with the UCAC3 values, with a mean offset
of -8.8 $\pm$ 5.5 mas/yr in RA and 1.1 $\pm$ 5.1 mas/yr in DEC (Figure
\ref{fig_compare_ucac}). We correct these biases to find the proper motions of
TWA sources without UCAC3 measurements, i.e. TWA 13AB, 15AB, 26, and 29.

\section{Results}

The results of our parallax survey are presented in Table
\ref{tab_astrometry}, including the relative proper motions and parallaxes
from the iterative solution, as well as the estimates of zero-point parallax
in each field and the resulting absolute parallaxes.

Two targets, TWA 13 and TWA 15, are visual binaries for which we
obtained independent astrometry for the two stars in each system.  Their
solutions agree within their uncertainties. 

\subsection{Notes on Individual Sources}

\subsubsection{TWA 5 \label{SecTWA5}}
\begin{figure}
\includegraphics[scale=0.75]{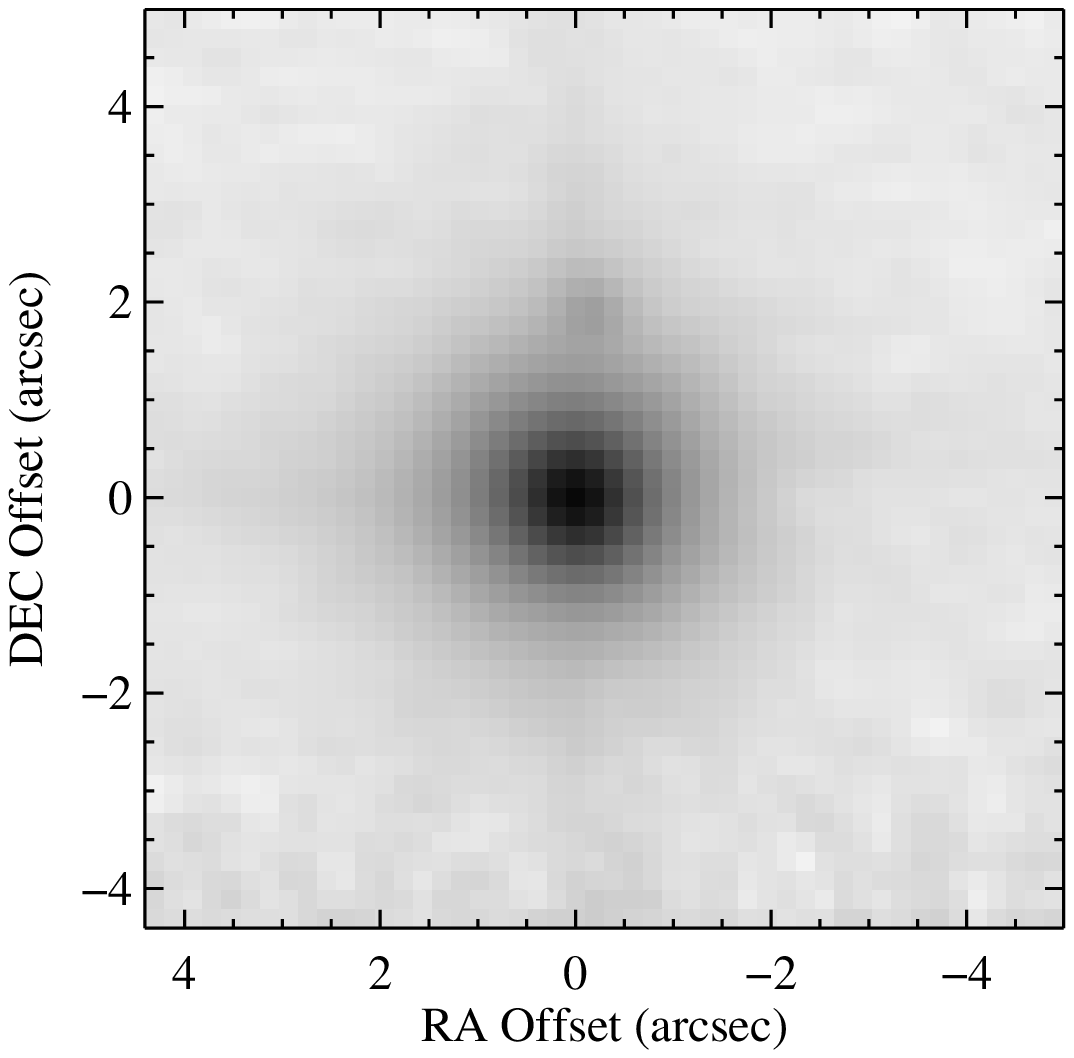}
\figcaption{Image of TWA 5A and 5B from the sum of 782 individual 0.2 s images
  taken 2009 April 11 that were shifted to stack the brightest pixel at the center. TWA 5B is
  visible 2\arcsec north of the primary. \label{fig_twa5}}
\end{figure}

TWA 5 (i.e. TWA 5A) has a companion brown dwarf (BD) TWA 5B
\citep{Webb:1999}. It is visible in CAPSCam images taken during good seeing
(Figure \ref{fig_twa5}), but the BD is faint enough and widely separated
enough that it does not contribute to the PSF centroiding of the primary star.

We measure the location of 5B in the 2009 April 11 and 2009
June 9 epochs, which both had excellent seeing of $\sim$0\farcs7. Because we
were integrating on TWA 5A in the GW at 0.2 s and the reference FF for 30~s or
12~s, respectively, we have 2250 (720) GW images to use for ``lucky imaging''
to select the best few hundred images on each date. For each epoch, we shift
and add these on the brightest pixel to form a final image with high quality
-- FWHM of 0\farcs54 on 11 April and 0\farcs60 on 9 June.  In these images, we
measure the separation and position angles of the brown dwarf 5B with respect
to 5A. The location of 5A is well-determined by the shift-and-add process.  To
find the TWA 5B centroid, we examine individually the flux in slices of each
row and column around its location. In each slice, we subtract a smooth
continuum from the bright star, and then fit the peak produced by 5B with a
Gaussian. We then average the individual slice locations weighted by the
height of the Gaussian in each slice to produce separate centroids in RA and
DEC. Unfortunately, TWA 5B falls nearly on top of a diffraction spike that
limits our ability to accurately centroid.  The average separation on the two
dates is 2\farcs00 $\pm$ 0\farcs10.

We use 2MASS sources in the larger astrometric frame to solve for any PA
offset between CAPSCam and the 2MASS reference frame and find offsets of 0.17
and 0.22 deg on the two dates. A match of the CAPSCam and 2MASS fields is also
used to measure the CAPSCam pixel scale of 0.1958 $\pm$ 0.0005
arcsec~pixel$^{-1}$. The average position angle of B with respect to A in the
2009 epochs is 358.3 $\pm$ 2.0 degrees. Both the separation and position angle
are consistent with the measurements of \citet{Neuhauser:2010} taken one year
earlier.

TWA 5A itself is a binary whose orbit of period 6 yr was measured with AO
imaging \citep{Konopacky:2007}. We determine its distance to be 50.1 $\pm$ 1.8
pc. The total dynamical mass of the components from the astrometric orbit
depends strongly on distance ($\propto$D$^3$), which was taken in Konopacky et
al. to be 44 $\pm$ 4 pc from the convergent point analysis by
\citet{Mamajek:2005}. Our parallactic distance is a 14\% change from the
Mamajek estimate, which makes a significant change in the inferred dynamical
mass. The mass is now estimated at 1.05 $\pm$ 0.22 M$_\odot$ instead of 0.71
$\pm$ 0.14 M$_\odot$.

When we correct the absolute magnitude of TWA 5A for binarity using
$\Delta$mag$\approx$1.1 at H-band as measured by
\citet{Konopacky:2007}, we find that this brings the dynamical mass into
good agreement with the \citet{Baraffe:1998} pre-main sequence tracks,
which predict a total mass of 1.2 M$_\odot$ and an age of 10 $\pm$ 1 Myr from
our interpolation in Section \ref{ages} (see also Figure 4 of Konopacky et
al.).  

This assumed that there were two and only two components in TWA 5A. However,
there have been suggestions of additional rapid, high velocity variability
that would suggest a separate spectroscopic binary within TWA 5Aa or 5Ab.
\citep{Torres:2003} report that TWA 5A may be double-lined with possible
day-scale velocity changes, but does not present individual RV
measurements. \citet{Reid:2003} find two components separated by 30 km
s$^{-1}$, and may have observed the AO binary near periastron. Periastrons
were in 1998.40, 2004.34, and 2010.28 according to the \citep{Konopacky:2007}
ephemeris.  \citet{Song:2002} report a single velocity of -30.6 $\pm$ 6.6 km
s$^{-1}$ measured in 2001-2002 at apastron.  Other measurements find mean
velocity of $\sim$13 km s$^{-1}$ with variability and line shapes consistent
with motion of a few km s$^{-1}$
\citep{Torres:2006,Torres:2008,Shkolnik:2011}.  TWA Aa and Ab are of very
similar spectral type and brightness, and have significant rotational velocity
$>$ 30 km s$^{-1}$ \citep{Torres:2003,Reid:2003}, so their lines should be
difficult to distinguish. The AO orbit predicts a maximum relative velocity
of $\sim$18 km s$^{-1}$. If there is a third star within TWA 5A, then the age
of TWA 5Aa/Ab from the \citep{Baraffe:1998} tracks would be underestimated
(i.e. they would have lower luminosity). Their masses from the tracks would be
essentially unchanged because the tracks are nearly vertical (see Figure
\ref{fig_hrcloseup}), but the addition of the mass of the third component
would imply that the tracks overestimate the dynamical masses of Aa/Ab
alone. 

\subsubsection{TWA 11C}

\citet{Kastner:2008} identified 2M1235 as a likely tertiary companion to HR
4796A, ie. as TWA 11C. Its parallactic distance of 69.0 $\pm$ 2.4 pc indeed
agrees well with the Hipparcos distance to HR 4796A of 72.8 pc $\pm$ 1.7 pc
\citep{vanleeuwen:2007}.

\subsubsection{TWA 16}

\begin{figure}
\includegraphics[scale=0.75]{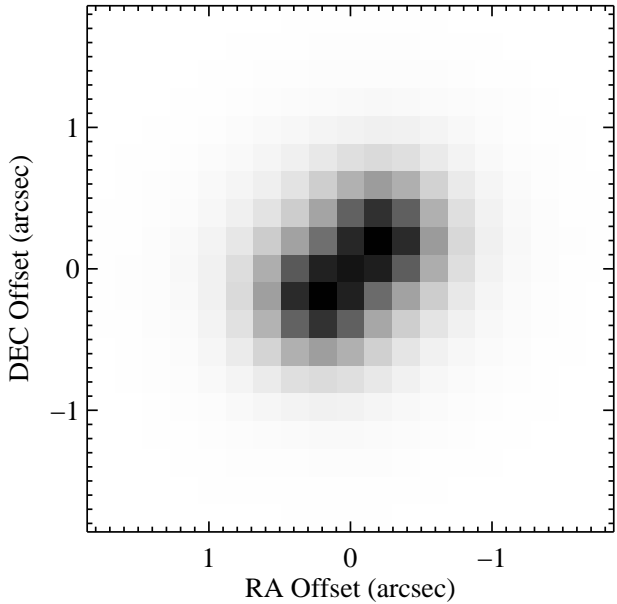}
\figcaption{Image of the close binary TWA 16 on 2009 June 8 with a separation of
  0\farcs61 $\pm$ 0\farcs02 and PA of 44$^\circ$.' \label{fig_twa16}}
\end{figure}

\citet{Zuckerman:2001} reported that TWA 16 was a close visual binary with
separation $\sim$0\farcs67 and flux ratio $\sim$0.9.  The CAPSCam images
reveal an elongated source in all epochs and resolve two sources during the
observations with the best seeing (Figure \ref{fig_twa16}). The pipeline does
not identify two sources there, however, so the measured parallax and proper
motion are for the photocenter of the system.  For the 2009 June 8 data, which
had the best seeing, we used ``lucky imaging'' to select the best few hundred
GW images.  We shift and added these on the brightest pixel to form a final
image with high quality and then used PSF fitting to measure the separation as
0\farcs61 $\pm$ 0\farcs02, PA as 44$^\circ$, and flux ratio as 0.97.

\section{Discussion}

\subsection{Pre-Main Sequence Track Ages \label{ages}}

We determine absolute magnitudes using our parallaxes and 2MASS apparent
magnitudes.  We also correct for binarity for six sources. TWA 2 is a visual
binary with $\Delta$mag$\approx$1 \citep{Webb:1999}. TWA 5A (see Section
\ref{SecTWA5}) is a close (speckle/AO resolved) binary with
$\Delta$mag$\approx$1.1 \citep{Konopacky:2007}.  TWA 14 is an approximately
equal brightness SB (\citet{Jayawardhana:2006} and Shkolnik, 2011, personal
communication). TWA 16 is a visual binary with a flux ratio of 0.9
\citep{Zuckerman:2001}. TWA 20 is a SB for which we assume the components are
equal brightness \citep{Jayawardhana:2006}. TWA 23 is a SB with equal
brightness components \citep{Shkolnik:2011}. No attempt has been made to
correct for extinction, which is small in the near-infrared due to the
closeness of the stars and the absence of edge-on optically thick
circumstellar disks in our sample. The tabulated uncertainties in absolute
magnitude include the photometric uncertainty in the 2MASS H-band measurement
and the parallax uncertainty.

We obtain effective temperatures by converting literature spectral types to
temperature using the intermediate scale of \citet{Luhman:1999} for the M-type
stars and tabulated values in \citet{Hartigan:1994} for the earlier type
stars. These are all given in Table \ref{tab_stellar}. Most of the spectral
types come from recent compilations that use TiO band or other spectral index
fitting and should be mutually consistent and good to $\sim$75~K. TWA 25 has
no published spectral type; to obtain its temperature, we fit a Kurucz model
to its photometry. TWA 29 (DEN1245) is the latest spectral type object for
which we measured a parallax and sits near the M-L transition. The
spectral-type to effective temperature conversion is not well known for such
objects, and we approximate it at 2250~K. Finally, we must note that
historical optical and new infrared spectral types for TW Hya do not agree
\citep{Webb:1999, Vacca:2011}. TW Hya's optical spectrum
has been typed as K7V, 4000~K, but the
\citet{Vacca:2011} determination of 3400~K is likely to be too cool
(N. Calvet, 2012, personal communication); we have chosen an intermediate
value of the T$\rm_{eff}$ of 3615 K.

\begin{figure}
\includegraphics[scale=0.95]{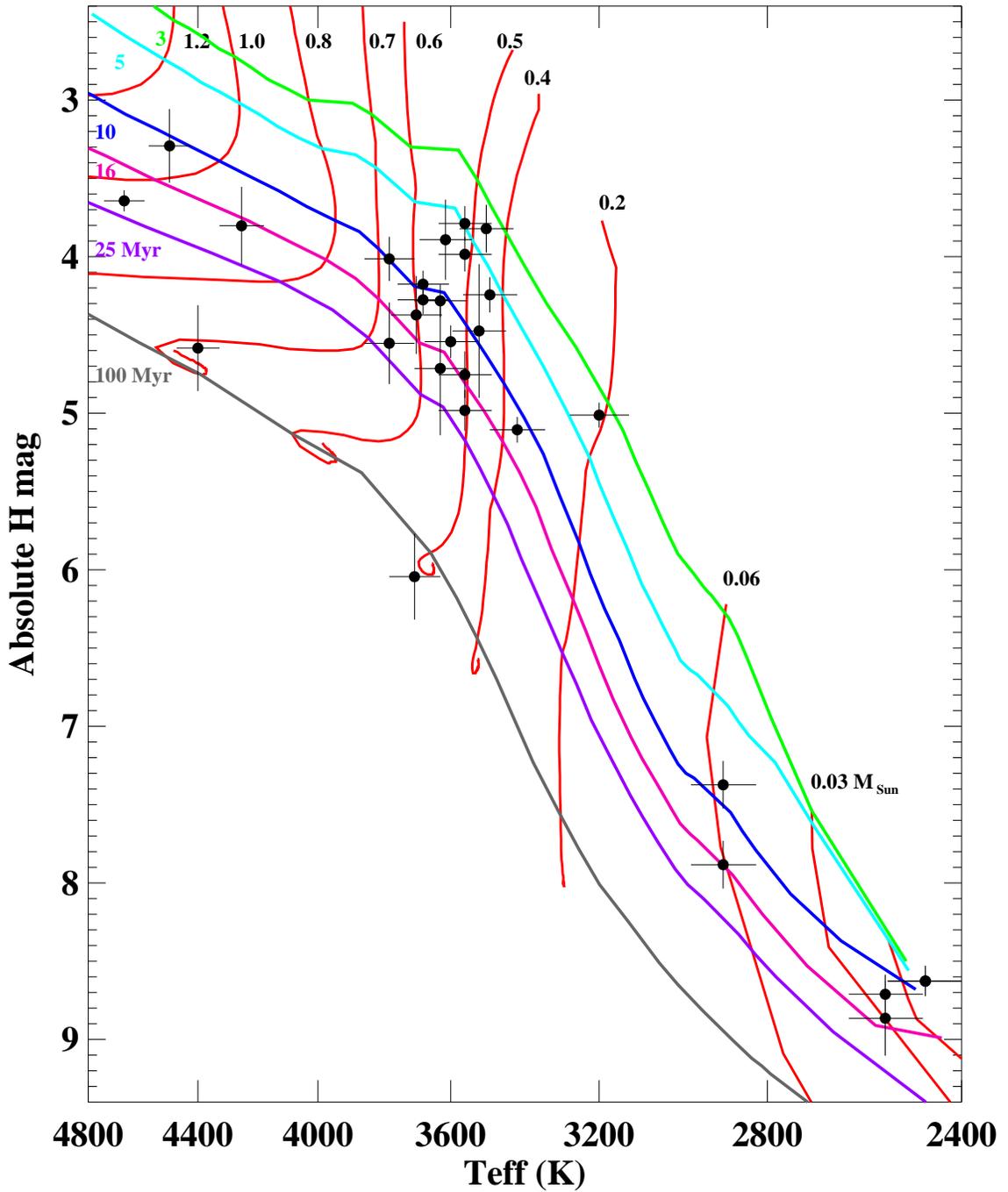}
\figcaption{TWA stars with parallaxes, and therefore absolute magnitudes,
  plotted on theoretical pre-main sequence tracks from \citet{Baraffe:1998} and
  \citet{Chabrier:2000}. Effective temperatures largely come from converting
  published spectral types to temperature and have uncertainties (as shown) of
  at $\sim$75 K (one half of a spectral type). The apparent scatter in ages is
  discussed in section \ref{ages} and Fig. \ref{fig_agehist}. \label{fig_HR}}
\end{figure}

\begin{figure}
\includegraphics[scale=0.95]{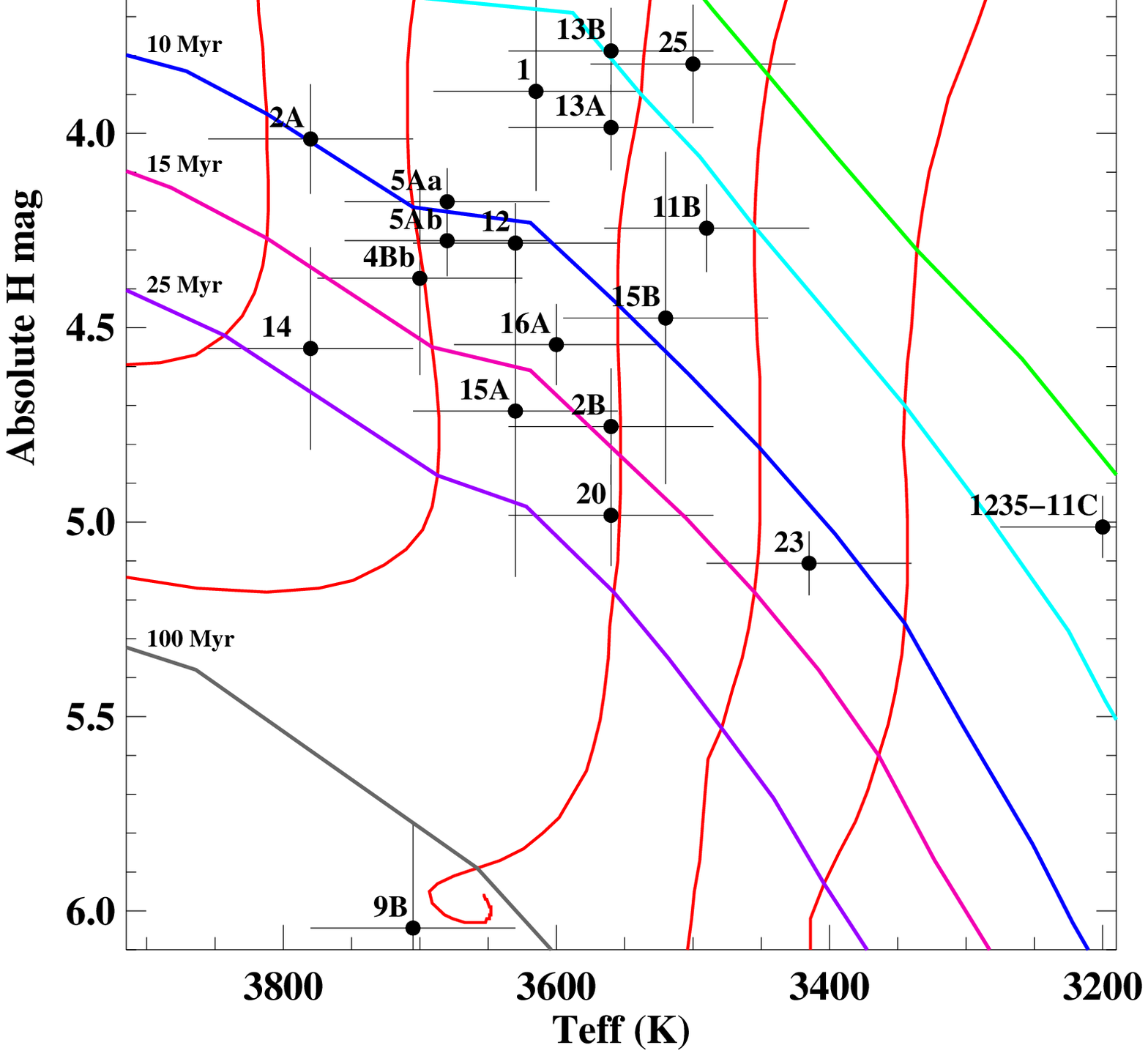}
\figcaption{Closeup of Figure \ref{fig_HR} in the region around
  T$\rm_{eff}$=3600~K where most of the stars sit. \label{fig_hrcloseup}}
\end{figure}

Theoretical isochrones overplotted with the data for all stars with parallaxes
(literature as well as this work) are shown in Figure \ref{fig_HR}.  We have
chosen the \citet{Baraffe:1998} tracks with Y=0.775, mixing length parameter 1
for m$<$0.6M$_\odot$ and Y=0.282, mixing length 1.9 for m$\geq$0.6M$_\odot$,
based on their relative success in reproducing multiple star coevality
\citep{White:1999}. The combination of the isochrones with different helium
abundances creates a small temperature discontinuity at 3500-3700 K, which is
unfortunately the temperature range of many of our stars (Table
\ref{tab_stellar}), but we interpolate over this region anyway.  We use the
DUSTY isochrones \citep{Chabrier:2000} for the stars with T$\rm_{eff}
<$2900~K.  We interpolate the combined theoretical tracks to estimate the ages
of all the stars at their nominal positions in absolute magnitude-T$_{\rm
  eff}$ space. The median age is 10.1 Myr. Individual ages are also given in
Table \ref{tab_stellar}, along with all the data plotted on the tracks.
Because of the abundance of stars with Teff$\sim$3600~K, Figure
\ref{fig_hrcloseup} shows a zoomed view of this part of the diagram.  From the
parallax uncertainty alone, the typical age uncertainty on each star is 3 Myr.
An age of 10 Myr is consistent with previous estimates, as summarized in
\citet{Fernandez:2008}.

\subsection{Galactic Space Motion and Membership}

For the analysis of the kinematics of the group, we use all stars with known
proper motions (from UCAC3 if available), parallaxes, and radial
velocities. These are given along with calculated UVWs in Table
\ref{tab_uvw}. Uncertainties in all three quantities are propagated into the
final velocity uncertainties. We also use positions from the 2MASS catalog
\citep{Skrutskie:2006}.  We then compute the mean velocity of the entire TW
Hya association.

The uncertainty weighted average velocities in Table \ref{tab_uvw} are [-10.1, -17.9, -8.0] $\pm$
[0.2, 0.2, 0.1] km~s$^{-1}$.  Two stars have velocities more than 3$\sigma$
from the mean of the association in at least two directions: TWA 9A and TWA
22. TWA 22 was already suspected not to be a member by \citet{Mamajek:2005}
and \citet{Teixeira:2009}, based on similar velocity arguments. However, TWA
9A is a ``classical'' member used to define the convergent point in the
Mamajek analysis. Given its discrepant age in Table \ref{tab_stellar}, as well
as its discrepant velocity, we conclude it is not a member or that its
Hipparcos distance is underestimated.  The new average velocities after
excluding these two stars are: [-10.9, -18.2, -5.3] $\pm$ [0.2, 0.2, 0.2]
km~s$^{-1}$. The standard deviation of the total velocities is 2.0 km~s$^{-1}$,
and the RMS deviation in the total velocity from the mean total velocity is
1.9 km~s$^{-1}$. These are considerably lower dispersions than obtained when
photometric distances are used \citep{Fernandez:2008}. 

The uncertainties on the mean velocities above are computed assuming the
stellar velocity distribution is Gaussian, i.e. standard deviation of
velocities divided by the square-root of the number of stars. However, a K-S
test reveals that the velocities are not Gaussianly distributed. Nor are they
uniformly distributed between their minimum and maximum values.  

To trace the stars back in time, we take their present positions and
three-dimensional space velocities and compute their locations in Galactic
coordinates for time-steps back every 100,000 years. To treat the distance and
velocity uncertainties properly, we do this in a Monte Carlo for 10000 trials
selecting each star's distance and velocity in each direction randomly in each
trial but distributed assuming the uncertainties for each individual star are
Gaussian.  Then, the centroid 3D location of the stars at each time is computed
as well as the average distance of the stars from this centroid.  The
time of best convergence is defined to be when the average distance is
minimized. We tested that the Galactic potential does not significantly affect
the motions over the short timescale of 10 Myr.

\begin{figure}
\plotone{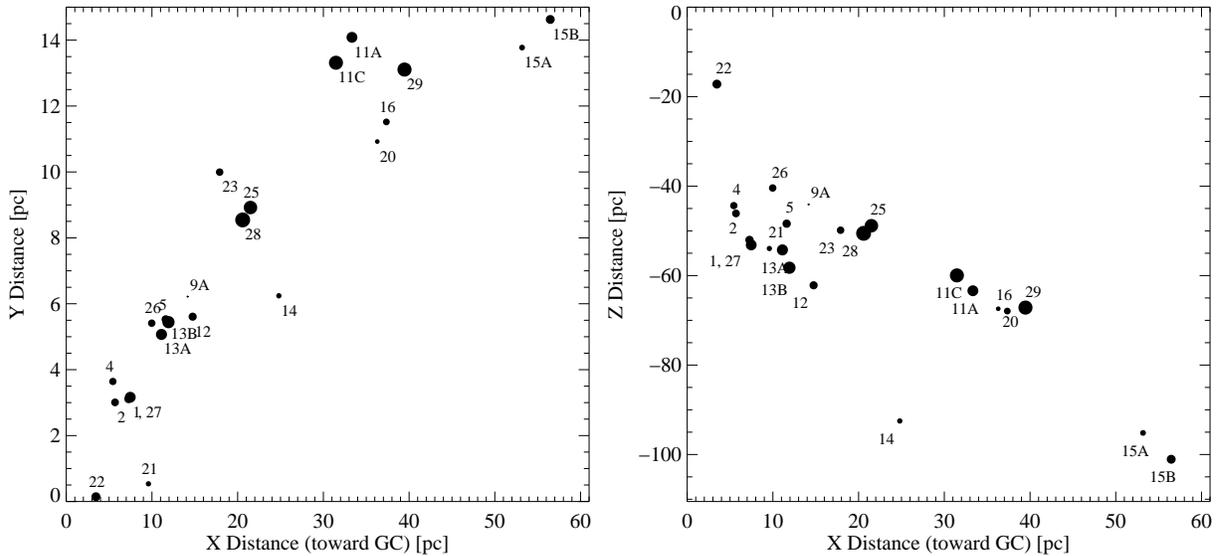}
\figcaption{Location of the TWA stars currently in Galactic coordinates. The
  size of the symbols is inversely proportional to age; that is, larger symbols
  indicate younger stars, as determined in
  Section \ref{ages}. There is no apparent correlation between age and
  location. \label{fig_location}}
\end{figure}

\begin{figure}
\includegraphics[scale=0.6]{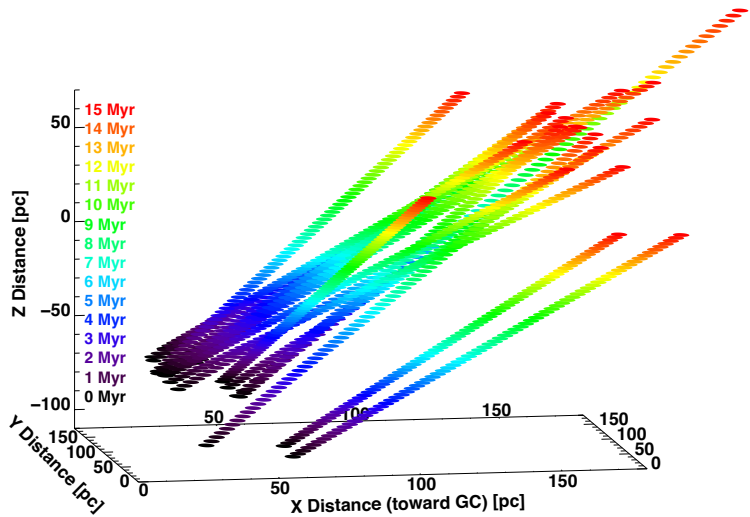}
\figcaption{Three-dimensional space motion of the TWA stars with measured
  parallaxes and radial velocities in Galactic coordinates. Black dots show
  the present locations of the stars and the colored points show the motion
  in 200,000 yr timesteps for 15 Myr. The parallel velocities mean that the stars are never much
  closer together than they are at present. This plot excludes TWA 9A and 22.\label{fig_distance}}
\end{figure}

The present average distance from their mean location for the stars in Table
\ref{tab_uvw}, excluding 9A and 22, is 20.4 pc.  The present-day locations of
the TWA stars with measured parallaxes and radial velocities is shown in
Figure \ref{fig_location}.  The velocities are nearly parallel, as shown in
Figure \ref{fig_distance}. The nominal closest approach of all the stars is 2
Myr ago, but has a mean distance of 19.2 pc from the center, meaning that
there is no time in the past when the stars are significantly more
concentrated than they are today. At the mean age of 10 Myr established in
section \ref{ages}, the mean distance from the center is 34 pc.

\begin{figure}
\includegraphics[scale=0.6]{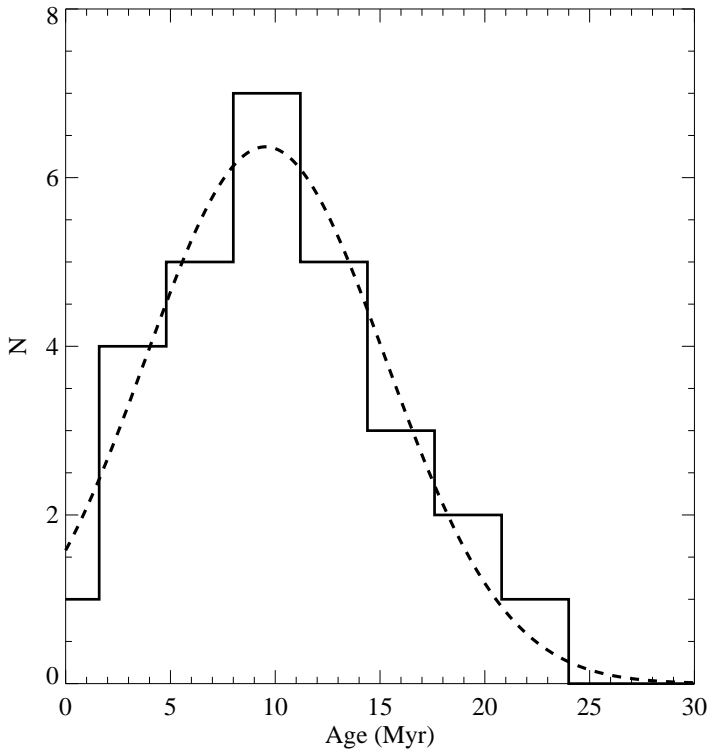}
\figcaption{Age histogram of the TWA stars with parallaxes, excluding 9AB and
  22, which have kinematics inconsistent with membership, and 29, which
  has a highly uncertain T$_{\rm eff}$ and therefore uncertain age. The bin size is 3
  Myr, which is the median uncertainty in the individual ages based on the
  uncertainties in their absolute magnitudes (i.e., parallaxes). The best fit
  Gaussian to the age distribution is overplotted with the dashed line and has
  a mean of 9 Myr and standard deviation of 6.8 Myr. The width of the age
  distribution exceeds what would be expected for a population of stars formed
  in a single burst. \label{fig_agehist}}
\end{figure}

Figure \ref{fig_agehist} displays a histogram of the ages excluding 9AB, 22,
and 29. There is a tail of stars to apparently larger ages while there
are no stars with inferred ages less than 3 Myr. The best Gaussian fit to the
age distribution has a mean of 9.5 Myr and standard deviation of 5.7 Myr. To
assess the reliability of these values for such a small sample, we repeated
the histogram fit in a Monte Carlo. In each trial, the age of each star was
drawn from a Gaussian distribution based on that individual's star mean age
and absolute magnitude uncertainty on the \citet{Baraffe:1998} tracks.  The
typical age uncertainty on each star is 3 Myr. The mean age over all the
trials was 8.7 Myr and the width of the age distribution over all the trials
was 6.5 Myr. Thus, the age histogram is robust against the individual age
uncertainties, and the width of the distribution compared to the typical age
uncertainty indicates that there is a real spread in derived ages. 

\section{Conclusions}

The identification of TWA members has largely been based on youth plus
similarity to the young star TW Hya in terms of location on the sky, proper
motion, and radial velocity.  Co-evality was thought to follow under the
assumption that these stars with similar motions formed from the same raw
material at the same time. With parallaxes to fourteen primary stars
identified as TWA members, we have greatly expanded the knowledge of the
kinematics of these young stars. We find that although they do share a common space
motion, the stars do not appear to have formed in a concentrated volume with a
well-defined expansion velocity. The TWA stars appear to have formed over a
larger volume than they presently occupy.

Nor do the stars appear to be completely co-eval, as the stars studied here
have ages that range from 3--23 Myr as derived from their locations on
pre-main sequence tracks. Although uncertainties in the distance and the
effective temperatures allow for several Myr of uncertainty in individual
ages, it would be extremely difficult to force them all to a common age.  This
apparent age spread could be due to a real difference in the times of
formation of the stars or it could be due to the lasting effects of episodic
accretion \citep{Baraffe:2009}.

The spatial distribution of these nominal TWA stars of 40--60 pc is largely
filamentary in nature, which naturally leads to some conclusions about their
provenance. Perhaps the stars formed in an extended wisp of molecular cloud,
probably one related to the Scorpius-Centaurus complex that is nearby on the
sky but 80 pc further away. The Galactic V and W velocities of TWA are very
similar to the older Upper Centaurus Lupus (UCL) and Lower Centaurus Crux
(LCC) subgroups of Sco-Cen \citep{Chen:2011} at ages of 15--17 Myr
\citep{Mamajek:2002}. TWA is located near to LCC in Galactic coordinates but
their separation in distance, age, and velocity distinguish the two groups.
\citet{Fernandez:2008} showed that TWA was $sim$45 pc from LCC 8 Myr ago and
could have been subjected to 0.5 supernovae per Myr. Multiple supernova shocks
could have triggered star formation in dense parts of the filamentary
progenitor TWA cloud over a few million years and not in a regular progression
from one side to the other.  \citep{Ortega:2009} also suggest that stellar
winds and supernovae from LCC and UCL could compress gas in the region of TWA,
although the mechanism for an extended time of star formation is less clear in
this case.

\acknowledgements

The Las Campanas Observatory staff and operators of the duPont telescope,
particularly Oscar Duhalde, Javier Fuentes, Herman Oliveras, Patricio Pinto,
and Andr\'es Rivera, made the CAPSCam observations smooth and efficient.
Rebecca Rattray made helpful analyses of portions of these TWA data during an
undergraduate internship at DTM in 2009. CAPSCam was built with support from
the NSF ATI program and Carnegie Institution of Washington.  We acknowledge
support for the observing by the NASA Astrobiology Institute under cooperative
agreement NNA09DA81A.  This work makes use of the Simbad database and Vizier
catalogue access tool, CDS, Strasbourg, France, and the 2MASS survey, which is
a joint project of the University of Massachusetts and the Infrared Processing
and Analysis Center/California Institute of Technology, funded by the National
Aeronautics and Space Administration and the National Science Foundation.

\bibliographystyle{apj}
\bibliography{twapaperlib2}

\begin{deluxetable}{lllllll}
\tablecaption{Observation Log\label{tab_obs}}
\tablewidth{0pt}
\tabletypesize{\scriptsize}
\rotate
\tablehead{
\colhead{Target} &\colhead{Other}
                 &\colhead{I} &\colhead{Ref} 
                                &\multicolumn{2}{c}{Integration Times}
&\colhead{Epochs of Observation}\\
\colhead{}       &\colhead{Name}  
                 &\colhead{(mag)}&\colhead{}
                                &\colhead{GW (s)} &\colhead{FF (s)} &\colhead{(JD)}
}
\startdata
TWA 2   &          &8.9    &1,2  &0.2  &40 & 2454167.7, 2454809.8, 2454818.8, 2454931.7, 2454989.5, 2454992.5, 2455222.8, 2455295.9, 2455297.9, 2455369.6 \\ 
TWA 5   &          &9.3      &5  &0.2  &12 & 2454664.5, 2454810.8, 2454932.7, 2454990.4, 2454991.5, 2455222.9, 2455368.7, 2455579.9\\
TWA 11C &2M1235-39 &11.2     &2  &--   &30 & 2454861.8, 2454929.6, 2454989.6, 2454992.5, 2455224.8, 2455369.7\\
TWA 12  &          &10.5     &5  &0.6  &20 & 2454167.7, 2454168.1, 2454663.4, 2454809.8, 2454859.8, 2454932.7, 2454991.5, 2455222.9\\
TWA 13  &        &10.1/10.1  &3  &0.5  &20 & 2454810.8, 2454859.8, 2454930.7, 2454988.5, 2455218.8\\
TWA 14  &          &10.7     &5  &0.5  &20 & 2454810.8, 2454859.7, 2454929.8, 2454986.5, 2455218.8\\
TWA 15  &        &11.8/11.9  &4  &1.0  &45 & 2454813.8, 2454930.7, 2454990.5, 2455219.8, 2455636.8\\
TWA 16  &          &10.2     &4  &1.0  &30 & 2454861.8, 2454929.7, 2454988.5, 2454989.5, 2454991.5, 2455219.9\\
TWA 20  &          &10.7     &5  &1.0  &30 & 2454859.8, 2454929.7, 2454987.5, 2455223.9\\
TWA 21  &          &9.0      &5  &0.2  &30 & 2454813.7, 2454929.6. 2454985.5, 2455216.8, 2455297.9, 2455370.6\\
TWA 23  &          &10.1     &5  &0.5  &60 & 2454168.7, 2454291.5, 2454661.5, 2454811.8, 2454852.8, 2454861.8, 2454989.5, 2454992.4\\
TWA 25  &          &9.5      &1  &0.3  &45 & 2454168.3, 2454291.5, 2454663.5, 2454861.8, 2454991.5, 2455299.1\\
TWA 26  &2M1139-31 &15.8     &5  &--   &30 & 2454809.1, 2454930.7, 2454985.5, 2455217.8, 2455295.9, 2455584.1\\
TWA 29  &DEN1245-44 &18.0    &5  &--   &60 & 2454861.7, 2454990.6, 2455217.9, 2455368.7, 2455410.1\\
\enddata
\tablenotetext{a}{I-band magnitudes are from (1) USNOB1.0 (Vizier I/284; \citep{Monet:2003}),
  (2) UCAC3 (Vizier I/315), (3) \citet{Reid:2003}, (4) \citet{Zuckerman:2001} or
  (5) DENIS (Vizier B/denis)}
\end{deluxetable}

\begin{deluxetable}{llllll}
\tablecaption{Astrometric Results\label{tab_astrometry}}
\tablewidth{0pt}
\tablehead{
\colhead{Target}
        &\colhead{$\pi_{rel}$}
                             &\colhead{$\mu_{RA\cos{DEC},rel}$}
                                             &\colhead{$\mu_{DEC,rel}$}
                                                                &\colhead{Zero-point}
                                                                                       &\colhead{$\pi_{abs}$}
}
\startdata
TWA 2         &21.76  $\pm$ 1.26   &-80.8  $\pm$  0.9    &-18.6  $\pm$  0.9     &  0.28 $\pm$ 0.30      & 21.48 $\pm$ 1.30\\
TWA 5         &20.07  $\pm$ 0.67   &-75.7  $\pm$  1.0    &-21.1  $\pm$  4.4     &  0.10 $\pm$ 0.19      & 19.97 $\pm$ 0.70\\
TWA 11C       &14.55  $\pm$ 0.38   &-45.0  $\pm$  0.8    &-26.1  $\pm$  1.2     &  0.06 $\pm$ 0.34      & 14.49 $\pm$ 0.51\\
TWA 12        &15.43  $\pm$ 0.59   &-54.4  $\pm$  1.1    &-16.4  $\pm$  1.4     & -0.16 $\pm$ 0.37      & 15.59 $\pm$ 0.70\\
TWA 13A (NW)  &17.89  $\pm$ 0.68   &-57.7  $\pm$  1.7    &-13.6  $\pm$  0.9     & -0.09 $\pm$ 0.23      & 17.98 $\pm$ 0.72\\
TWA 13B (SE)  &16.66  $\pm$ 0.70   &-59.3  $\pm$  2.6    &-12.2  $\pm$  2.1     & -0.09 $\pm$ 0.23      & 16.75 $\pm$ 0.74\\
TWA 14        &10.15  $\pm$ 1.19   &-36.3  $\pm$  2.7    & -4.4  $\pm$  3.3     & -0.27 $\pm$ 0.21      & 10.42 $\pm$ 1.21\\
TWA 15A (NE)  & 8.27  $\pm$ 1.61   &-28.8  $\pm$  1.6    &-11.5  $\pm$  1.2     & -0.30 $\pm$ 0.13      &  8.57 $\pm$ 1.62\\
TWA 15B (SW)  & 8.80  $\pm$ 1.72   &-27.8  $\pm$  2.3    &-11.0  $\pm$  2.3     & -0.30 $\pm$ 0.13      &  9.10 $\pm$ 1.72\\
TWA 16        &13.04  $\pm$ 0.49   &-41.4  $\pm$  1.7    &-26.9  $\pm$  4.3     &  0.28 $\pm$ 0.12      & 12.76 $\pm$ 0.50\\
TWA 20        &12.85  $\pm$ 0.59   &-44.3  $\pm$  1.2    &-22.9  $\pm$  2.5     & -0.08 $\pm$ 0.15      & 12.93 $\pm$ 0.61\\
TWA 21        &18.20  $\pm$ 0.46   &-56.4  $\pm$  0.9    &  4.8  $\pm$  1.2     & -0.05 $\pm$ 0.17      & 18.25 $\pm$ 0.49\\
TWA 23        &18.41  $\pm$ 0.33   &-63.8  $\pm$  0.6    &-27.2  $\pm$  1.5     & -0.14 $\pm$ 0.35      & 18.55 $\pm$ 0.48\\
TWA 25        &18.39  $\pm$ 1.23   &-68.7  $\pm$  1.2    &-28.3  $\pm$  1.2     & -0.09 $\pm$ 0.14      & 18.48 $\pm$ 1.24\\
TWA 26        &23.38  $\pm$ 2.54   &-81.2  $\pm$  3.9    &-27.7  $\pm$  2.1     & -0.44 $\pm$ 0.46      & 23.82 $\pm$ 2.58\\
TWA 29        &12.61  $\pm$ 2.06   &-40.3  $\pm$ 11.7    &-20.3  $\pm$ 17.0     & -0.05 $\pm$ 0.18      & 12.66 $\pm$ 2.07 \\
\enddata
\end{deluxetable}

\begin{deluxetable}{lcllllr}
\tablecaption{Stellar Parameters\label{tab_stellar}}
\tabletypesize{\scriptsize}
\tablewidth{0pt}
\tablehead{
\colhead{Star}  
         &\colhead{Sp. Type} &\colhead{Sp. Type Ref} &\colhead{Teff}
                        &\colhead{H(abs)} &\colhead{H unc}
                                         &\colhead{BCAH98 age}}
\startdata
1         &K7.0  &\citet{Webb:1999}       &3615  &3.91  &0.26 &6\\     
2A        &M0.5  &\citet{Webb:1999}       &3780  &4.03  &0.14 &9\\     
2B        &M2    &\citet{Webb:1999}       &3560  &4.77  &0.15 &17\\    
4Aab      &K4.0  &\citet{Prato:2001}      &4500  &3.31  &0.24 &11\\    
4Ba       &K4.0  &\citet{Soderblom:1998}  &4250  &3.82  &0.25 &15\\    
4Bb       &K4.0  &\citet{Soderblom:1998}  &3700  &4.39  &0.25 &12\\    
5A        &M1.5  &\citet{Konopacky:2007}  &3680  &4.19  &0.09 &9\\     
5B        &M8.5  &\citet{Konopacky:2007}  &3680  &4.29  &0.09 &11\\     
9A        &K5.0  &\citet{Webb:1999}       &4400  &4.60  &0.27 &63\\    
9B        &M1.0  &\citet{Webb:1999}       &3705  &6.06  &0.27 &150\\   
11B       &M2.5  &\citet{Webb:1999}       &3490  &4.26  &0.11 &6\\     
11C       &M4.5  &\citet{Kastner:2008}    &3200  &5.03  &0.08 &4\\     
12        &M1.6  &\citet{Shkolnik:2011}   &3630  &4.30  &0.10 &10\\    
13A       &M2.0  &\citet{Sterzik:1999}    &3560  &4.00  &0.11 &6\\     
13B       &M2.0  &\citet{Sterzik:1999}    &3560  &3.80  &0.11 &5\\     
14        &M0.6  &\citet{Shkolnik:2011}   &3780  &4.57  &0.26 &19\\    
15A       &M1.5  &\citet{Zuckerman:2001}  &3630  &4.73  &0.43 &17\\    
15B       &M2.2  &\citet{Shkolnik:2011}   &3520  &4.49  &0.43 &9\\     
16A       &M1.8  &\citet{Shkolnik:2011}   &3600  &4.56  &0.10 &14\\    
20        &M2.0  &\citet{Reid:2003}       &3560  &5.00  &0.13 &23\\    
21        &K3.5  &\citet{Zuckerman:2004}  &4665  &3.66  &0.07 &19\\    
22A       &M6.0  &\citet{Bonnefoy:2009}   &2900  &7.39  &0.15 &6\\     
22B       &M6.0  &\citet{Bonnefoy:2009}   &2900  &7.90  &0.15 &12\\    
23        &M2.9  &\citet{Shkolnik:2011}   &3415  &5.12  &0.08 &12\\    
25        &M2.0  &color                   &3500  &3.84  &0.15 &4\\     
26        &M8.0  &\citet{Mohanty:2003}    &2550  &8.88  &0.24 &12\\    
27        &M8.0  &\citet{Mohanty:2007}    &2550  &8.73  &0.12 &10\\    
28        &M8.5  &\citet{Scholz:2005}     &2470  &8.64  &0.06 &3\\     
29        &M9.5  &\citet{Looper:2007}     &2250  &9.31  &0.37 &4\\     
\enddata
\end{deluxetable}

\begin{deluxetable}{lcccccccccccccc}
\tablecaption{Velocities\label{tab_uvw}}
\tabletypesize{\scriptsize}
\rotate
\tablewidth{0pt}
\tablehead{
\colhead{Star}
&\colhead{$\pi$} &\colhead{$\sigma\pi$}
&\colhead{$\mu_{RA\cos{DEC}}$} &\colhead{$\mu_{DEC}$} 
&\colhead{$\sigma\mu_{RA}$} &\colhead{$\sigma\mu_{DEC}$} 
&\colhead{RV}&\colhead{$\sigma$RV}
&\colhead{U}&\colhead{$\sigma$U}
&\colhead{V}&\colhead{$\sigma$V}
&\colhead{W}&\colhead{$\sigma$W}\\
\colhead{} 
&\colhead{(mas)} &\colhead{(mas)}
&\colhead{(mas}&\colhead{(mas}&\colhead{(mas}&\colhead{(mas}
&\colhead{(km} &\colhead{(km}
&\colhead{(km} &\colhead{(km}
&\colhead{(km} &\colhead{(km}
&\colhead{(km}&\colhead{(km}\\
\colhead{} 
&\colhead{} &\colhead{}
&\colhead{yr$^{-1}$)}&\colhead{yr$^{-1}$)}&\colhead{yr$^{-1}$)}&\colhead{yr$^{-1}$)}
&\colhead{s$^{-1}$)} &\colhead{s$^{-1}$)}
&\colhead{s$^{-1}$)} &\colhead{s$^{-1}$)}
&\colhead{s$^{-1}$)} &\colhead{s$^{-1}$)}
&\colhead{s$^{-1}$)}&\colhead{ s$^{-1}$)}
}
\startdata
1       &18.6\tablenotemark{a} &2.1    & -70.2 &-13.7  &2.5 &1.1  &  12.7 \tablenotemark{d} &  0.2  &-11.9 &1.8   &-18.0 &0.7  &-5.2 &1.0\\
2       &21.5                  &1.3    & -91.1 &-21.0  &0.8 &0.8  &  11.0 \tablenotemark{d} &  0.1  &-13.8 &1.1   &-17.8 &0.4  &-6.3 &0.5\\
4       &22.3\tablenotemark{a} &2.3    & -91.7 &-28.2  &1.5 &2.4  &   9.2 \tablenotemark{d} &  1.0  &-13.0 &1.8   &-17.2 &0.9  &-6.0 &1.0\\
5       &20.0                  &0.7    & -82.6 &-22.6  &0.8 &1.0  &  13.3 \tablenotemark{e} &  2.0  &-11.8 &0.8   &-20.7 &1.8  &-4.8 &0.9\\
9A      &21.4\tablenotemark{a} &2.5    & -53.1 &-20.0  &1.9 &3.4  &   9.5 \tablenotemark{d} &  0.4  & -5.7 &1.3   &-14.4 &0.7  &-2.9 &0.9\\
11A     &13.7\tablenotemark{a} &0.3    & -53.3 &-21.2  &3.0 &4.0  &   6.9 \tablenotemark{f} &  1.0  &-10.8 &1.1   &-17.3 &1.1  &-5.2 &1.3\\
11C     &14.5                  &0.5    & -45.1 &-20.1  &2.4 &2.3  &   9   \tablenotemark{g} &  1.0  & -6.8 &0.9   &-16.8 &1.0  &-3.4 &0.8\\
12      &15.6                  &0.7    & -68.3 &-12.1  &2.7 &1.5  &  13.1 \tablenotemark{e} &  1.6  &-13.4 &1.2   &-20.1 &1.5  &-5.6 &0.8\\
13A     &18.0                  &0.7    & -66.4 &-12.5  &2.4 &1.8  &  11.7 \tablenotemark{d} &  0.6  &-11.4 &0.9   &-17.6 &0.6  &-3.9 &0.6\\
13B     &16.8                  &0.7    & -68.0 &-11.0  &3.1 &2.7  &  12.6 \tablenotemark{d} &  0.5  &-12.8 &1.1   &-18.9 &0.7  &-4.0 &0.8\\
14      &10.4                  &1.2    & -44.1 & -8.1  &1.4 &1.3  &  15.8 \tablenotemark{e} &  2.0  &-11.7 &2.2   &-21.9 &2.0  &-6.8 &1.2\\
15A     & 9.1                  &1.7    & -37.5 &-10.4  &2.4 &2.0  &  11.2 \tablenotemark{h} &  2.0  &-10.3 &3.6   &-20.4 &2.7  &-3.7 &1.5\\
15B     & 8.6                  &1.6    & -36.5 & -9.9  &2.9 &2.8  &  10.0 \tablenotemark{e} &  1.7  &-11.4 &3.8   &-19.8 &2.6  &-4.1 &1.9\\
16      &12.8                  &0.5    & -49.2 &-21.2  &1.6 &0.8  &   9.0 \tablenotemark{e} &  0.4  & -9.7 &0.8   &-18.6 &0.6  &-6.0 &0.4\\
20      &12.9                  &0.6    & -64.4 &-28.6  &3.1 &1.0  &   8.1 \tablenotemark{h} &  4.0  &-14.2 &2.3   &-21.0 &3.4  &-9.5 &1.3\\
21      &18.2                  &0.5    & -61.9 & 15.0  &1.2 &1.4  &  17.5 \tablenotemark{i} &  0.8  &-12.0 &0.5   &-20.2 &0.8  &-4.9 &0.4\\
22      &57.0\tablenotemark{b} &0.7    &-175.8 &-21.3  &0.8 &0.8  &  14.8 \tablenotemark{b} &  2.1  & -8.0 &0.4   &-17.1 &2.0  &-9.0 &0.1\\
23      &18.6                  &0.5    & -72.7 &-29.3  &0.9 &0.9  &   8.5 \tablenotemark{e} &  1.2  &-10.6 &0.6   &-18.2 &1.0  &-5.4 &0.6\\
25      &18.5                  &1.2    & -74.0 &-27.7  &0.8 &0.8  &   9.2 \tablenotemark{i} &  2.1  &-10.7 &1.4   &-18.7 &1.9  &-5.6 &1.0\\
26      &23.8                  &2.6    & -89.9 &-26.5  &4.2 &2.6  &  11.6 \tablenotemark{j} &  2.0  &-10.7 &2.0   &-18.8 &1.9  &-3.8 &1.3\\
27      &19.0\tablenotemark{c} &0.4    & -62.7 &-22.8  &1.7 &2.8  &  11.2 \tablenotemark{j} &  2.0  & -9.1 &0.6   &-16.6 &1.8  &-6.7 &1.0\\
\enddata
\tablecomments{Parallaxes are from this work unless otherwise noted in footnotes. Sources
    of RVs are given in footnotes.}
\tablenotetext{a}{Hipparcos -- \citet{vanleeuwen:2007}}
\tablenotetext{b}{\citet{Teixeira:2008}}
\tablenotetext{c}{Weighted average of \citet{Biller:2007}, \citet{Gizis:2007},
  and \citet{Ducourant:2008}}
\tablenotetext{d}{\citet{Torres:2003}}
\tablenotetext{e}{\citet{Shkolnik:2011}}
\tablenotetext{f}{Bright Star Catalog V/50}
\tablenotetext{g}{Assume same RV as for HR 4796B from \citet{Stauffer:1995}}
\tablenotetext{h}{\citet{Reid:2003}}
\tablenotetext{i}{\citet{Song:2003}}
\tablenotetext{i}{\citet{Mohanty:2003}}

\end{deluxetable}
\end{document}